# Flexible broadband polarization converter based on metasurface at microwave band*

Qi Wang(王奇)[1], Xiangkun Kong(孔祥鲲)[1,†], Xiangxi Yan(严祥熙)[1], Yan Xu(徐岩)[1], Shaobin Liu(刘少斌)[1], Jinjun Mo(莫锦军)[2], and Xiaochun Liu(刘晓春)[3]

[1] Key Laboratory of Radar Imaging and Microwave Photonics, Ministry of Education, College of Electronic and Information Engineering, Nanjing University of Aeronautics and Astronautics, Nanjing 211106, China
[2] School of Aeronautics and Astronautics, Central South University, Changsha 410083, China
[3] The Research Institute for Special Structures of Aeronautical Composite AVIC, The Aeronautical Science Key Lab for High Performance Electromagnetic Windows, Ji'nan 250000, China

**ABSTRACT:**

This paper proposes a flexible broadband linear polarization converter based on metasurface operating at microwave band. In order to achieve bandwidth extension property, long and short metallic arc wires, as well as the metallic disks placed over a ground plane, are combined into the polarizer, which can generate three neighboring resonances. Due to the combination of the first two resonances and optimized size and thickness of the unit cell, the polarization converter can have a weak incident angle dependence. Both simulated and measured results confirm that the average polarization conversion ratio is over 85% from 11.3 to 20.2 GHz within a broad incident angle from 0° to 45°. Moreover, the proposed polarization converter based on flexible substrates can be applied for conformal design. The simulation and experiment results demonstrate that our designed polarizer still keeps high polarization conversion efficiency even when it adheres on convex cylindrical surfaces. The periodic metallic structure of the designed polarizer has great potential application values in the microwave, terahertz and optic regimes.

* Project supported by 'the Fundamental Research Funds for the Central Universities' (kfjj20180401), Chinese Natural Science Foundation (Grant No.61471368), Aeronautical Science Foundation of China (20161852016), China Postdoctoral Science Foundation (Grant No. 2016M601802) and Jiangsu Planned Projects for Postdoctoral Research Funds (Grant No. 1601009B).

† Corresponding author. E-mail: xkkong@nuaa.edu.cn

# 1. INTRODUCTION:

Polarization is an important characteristic of electromagnetic(EM) wave and fully manipulating the polarization state is highly desirable in several electromagnetic applications [1,2]. Metamaterials, which have extraordinary electromagnetic properties that natural materials do not have [3,4], provide a very effective way to manipulate several basic properties of the EM wave and can be applied to polarizers [5,6,7], lenses [8,9], high gain antennas [10,11] and invisible cloak [12,13]. Compared with the conventional polarizers using the Faraday effect [14] and twisted nematic liquid crystal, the polarization converters based on the metamaterials are relatively thin (subwavelength thickness) and light. On the basis of metamaterial technology, many studies have been devoted to controlling the polarization state of EM waves recently. Many polarization converters based on anisotropic and chiral metamaterials have been proposed [15-17]. But these polarizers suffer from narrow bandwidth which extremely limits their practical applications. In order to expand the bandwidth, stacked multilayer structures [18-24], gold helix structures [25] and anisotropic high-impedance surfaces [26,27] can be utilized. However, these polarizers are hard to fabricate and integrate into existing systems due to their complex structures or unacceptable thickness. Some planar polarizers based on metasurfaces which are ultra-thin layered two-dimension metamaterials [27-35] were designed to achieve both wideband bandwidth and higher conversion efficiency. But, these proposed polarization devices may not keep high polarization efficiency within a broad incident angle. Thus, polarization devices with high polarization efficiency, wideband and high angular tolerance are highly desirable.

Rendering flexibility to metamaterials can make it possible to adhere flexible devices to all surfaces or wrap transparent and light metamaterials around objects, which many terahertz and microwave applications [36-45] can significantly benefit from. Kamali et al. [39] showed that cylindrical lenses covered with metasurfaces could have a function such as aspherical lenses focusing light to a point by using a flexible substrate. Youn et al. [45] proposed and demonstrated an optically transparent and flexible microwave absorber which provided a method to 3-D design. So flexible metamaterials open up a whole new range of practical applications. Additionally, to the best of our knowledge, there is a lack of polarizer based on flexible metamaterials or metasurfaces at microwave band.

In this paper, we design a reflective broadband linear polarizer composed of metallic disks and asymmetric arc metallic wires placed over flexible substrates. The long arc metallic wire resonator provides the main resonances. The short arc metallic wire and the metallic disk work as couplers to adjust the resonance frequencies and intensities. Both simulated and experimental results have demonstrated that the designed polarizer has wide operating bandwidth from 11.3GHz to 20.2GHz under normal incidence of both x- and y-polarized waves. Besides, it can keep high-efficient polarization conversion when the incident angle increases from 0° to 45°. The designed polarization device can also be applied for conformal design due to flexible substrates. The simulation and experiment results have demonstrated that our proposed polarizer can still keep high polarization conversion efficiency even adhered on convex cylindrical surfaces. The proposed periodical metallic structure has great potential application values in the microwave, terahertz and optic regimes.

## 2. STRUCTURE DESIGN AND THEORETICAL ANALYSIS

Fig.1 shows the structure of the polarization converter used in the simulations and experiments. The polarizer is composed of four layers with the periodic metallic structure on the top, the metallic ground plane on the bottom separated by two flexible material layers. The 30μm thick copper film is used as the metallic parts and polyethylene terephthalate(PET) is chosen as the flexible material with the thickness d1=0.12 mm. The relative dielectric constant of PET is 2.86 with a dielectric loss tangent of 0.053. Because the thickness of PET is not enough for the microwave polarizer design, polyvinyl chloride(PVC, $\varepsilon_r = 3(1 + j0.014)$ ) with thickness d2=2 mm is filled in the area between the PET layer and the ground plane. The top metallic unit cell is composed of three parts, the long arc metallic wire, the short arc metallic wire and the metallic disk. As shown in Fig. 2(a), the final geometrical parameters of the unit cell are: w=0.4mm, r1=r2=2.4mm, r3=1.5mm, α=190°, β=33°. The designed polarizer consists of 48×48 square unit cells (5mm ×5mm). The top metallic structure can be regarded as an anisotropic metamaterial layer for its asymmetrical structure along the x-axis and y-axis.

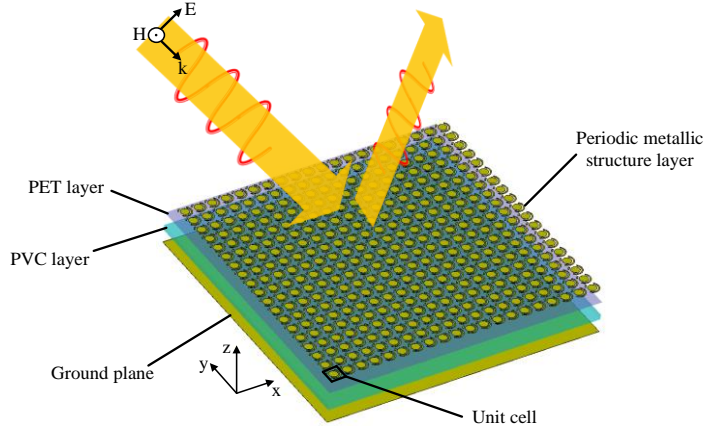

Fig.1. The structure diagram of the proposed polarization converter.

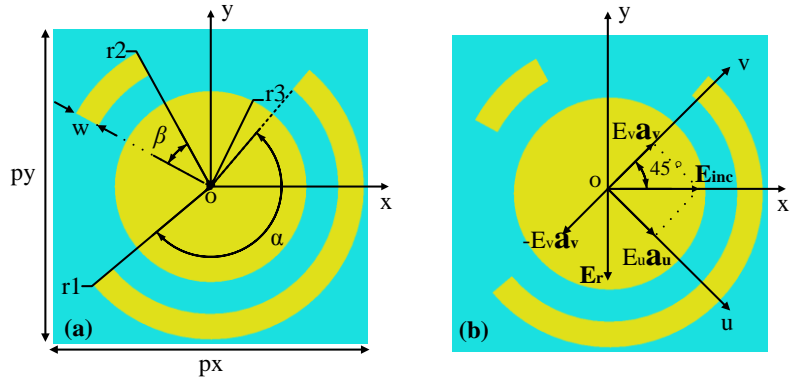

Fig.2. The unit cell of the metasurface: (a) Front view of the unit cell, (b) Intuitive scheme of x- to y-polarization EM wave conversion of the metasurface.

To illuminate the polarization conversion theory, we assume that an x-polarized electromagnetic wave is normally incident on the polarization converter (the same as a y-polarized wave). The electronic field of the incident electromagnetic wave denoted as $\mathbf{E}_{inc} = E_x \mathbf{a_x}$, can be decomposed into the u-polarized and v-polarized components denoted as $E_u \mathbf{a_u}$ & $E_v \mathbf{a_v}$, respectively, as shown in Fig. 2(b). In this way, the electric field vector $\mathbf{E}$ in the incident wave and the reflected wave can be expressed as equations (1) and (2), respectively.

$$\mathbf{E}_{inc} = E_x \mathbf{a_x} = E_x \cos(45°)(\mathbf{a_u} + \mathbf{a_v}) = E_u \mathbf{a_u} + E_v \mathbf{a_v} , \tag{1}$$

$$\mathbf{E_r} = \mathrm{R} \cdot \mathbf{E}_{inc} = (\mathbf{a_u}\ \mathbf{a_v}) \begin{pmatrix} r_{uu} & r_{uv} \\ r_{vu} & r_{vv} \end{pmatrix} \begin{pmatrix} E_u \\ E_v \end{pmatrix} , \tag{2}$$

where $\mathbf{a_u}$ and $\mathbf{a_v}$ are the unit vectors in u-axis and v-axis directions, $r_{uu}$ or $r_{vv}$ represents the co-polarization reflection coefficient and $r_{uv}$ or $r_{vu}$ represents the cross-polarization reflection coefficient, respectively, at the u-polarized or v-polarized wave incidence. Due to the symmetric structure along the u-axis, the amplitudes of $r_{uv}$ & $r_{vu}$ are near to 0, which means no cross-polarized reflection exists at u-polarized and v-polarized EM waves incidences. According to these conditions, equation (2) can be written as

$$\mathbf{E_r} = E_u r_{uu} \mathbf{a_u} + E_v r_{vv} \mathbf{a_v} = E_x \cos(45°)(r_{uu}\mathbf{a_u} + r_{vv}\mathbf{a_v}) , \qquad (3)$$

where $r_{uu} = |r_{uu}| e^{j\varphi_u}$ and $r_{vv} = |r_{vv}| e^{j\varphi_v}$.

Then, if conditions

$$|r_{uu}| = |r_{vv}| = 1, \qquad (4)$$

$$\Delta\varphi = \varphi_u - \varphi_v = \pi + 2n\pi , \qquad (5)$$

are satisfied (n is an integer), then

$$\mathbf{E_r} = E_x \cos(45°)e^{j\varphi_v}(\mathbf{a_u} + e^{j\Delta\varphi}\mathbf{a_v}) = E_x \cos(45°)e^{j\varphi_v}(\mathbf{a_u} - \mathbf{a_v}) = -E_x e^{j\varphi_v}\mathbf{a_y} , \qquad (6)$$

the polarization direction of the reflected light is transformed to the y-axis.

Fig.3 presents the simulated magnitudes and phases of $r_{uu}$ and $r_{vv}$ of the optimized polarizer based on CST Microwave Studio. It's clear that the magnitudes of $r_{uu}$ and $r_{vv}$ are near to 1 in the Fig.3.(a) and the absolute value of phase difference $\Delta\varphi$ is near to 180° from 11.3GHz to 20.2GHz in the blue square in Fig.3(b). So the conditions (4) and (5) are met, which means that the polarization direction of reflected wave is rotated by 90° with the proposed polarizer. From the view of the resonance [33], the electric field along the v-axis can excite two magnetic resonances (I and III), and along the u-axis can excite one electric resonance(II). Here the electric(magnetic) resonance does not necessarily mean out-of-phase (in-phase) reflection in [27] and the resonance type is judged through surface current distributions at the corresponding resonant frequency. Given the length of this paper, the current distributions at the three resonant frequencies are not presented here. In fact, the unit cell of the designed polarizer is a combined structure and it can excite multiple resonances. Moreover, main resonances are excited by the long arc metallic wire and the metallic disk because the resonance frequency of the short arc wire is out of the designed frequency band.

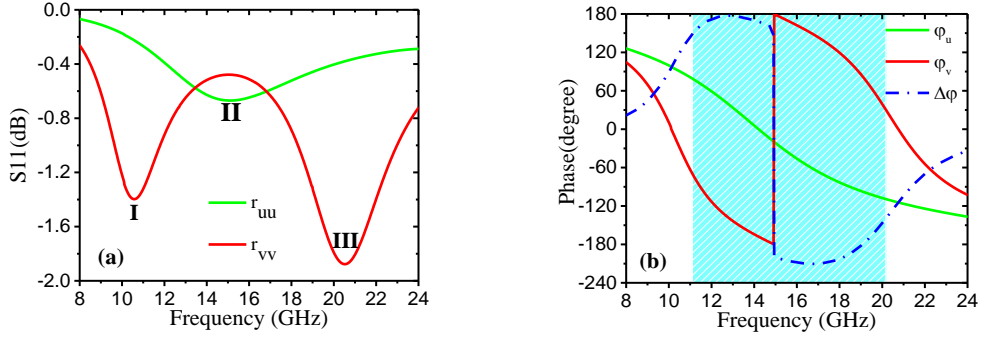

Fig.3. The co-polarization reflection coefficients for u- and v-axes. (a) Magnitude, (b) Phase and phase difference.

## 3. SIMULATION RESULTS

In order to verify the wideband property of the designed metasurface, full-wave simulations are performed with frequency domain solver in CST Microwave Studio. Set periodic boundary conditions in x and y directions and open conditions along the z-axis. The polarization conversion ratio (PCR) can be defined as

$$PCR = r_{yx}^2 / (r_{yx}^2 + r_{xx}^2) \ or \ r_{xy}^2 / (r_{xy}^2 + r_{yy}^2) \ , \tag{7}$$

$r_{xx}$ or $r_{yy}$ is the co-polarization reflection coefficient and $r_{yx}$ or $r_{xy}$ is the cross-polarization coefficient when x-polarized or y-polarized EM wave propagates along the z-axis direction. Because of the symmetric structure along u-axis, PCR, co- and cross-polarization reflection coefficients are same under the normal incidence of x- or y-polarized EM wave. We can find $r_{xy}$ & $r_{yx}$ are higher than -1 dB and $r_{yy}$ & $r_{xx}$ are lower than -10 dB in the frequency range from 11.3GHz to 20.2GHz in the Fig.4(a). It also can be seen from Fig.4(a) that there are two resonant frequencies, one at 13.0GHz and the other at 19.2GHz. As shown in Fig 4(b), PCR is higher than 90% from 11.3GHz to 20.2GHz under normal incidence. At the two resonant frequencies of 13.0GHz and 19.2GHz, the values of PCR are near to 100%. The simulation results demonstrate that the proposed polarizer can achieve wideband linear polarization conversion.

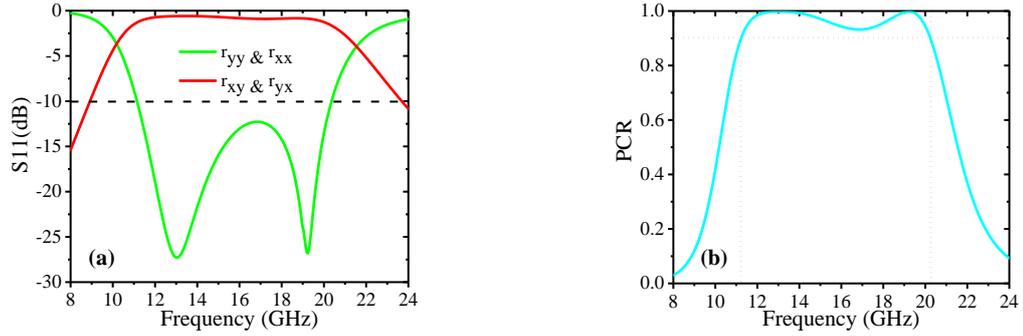

Fig.4. Under normal incidence, (a) simulated reflection coefficients, (b) PCR with r3=1.5mm.

### 3.1 Impact of metallic disk radius

To investigate the effects of the metallic disk on polarization conversion, we have carried out numerical simulations by changing the radius of the metallic disk. Fig.5(a) indicates that r3 can have great effects on the resonant frequencies and operating bandwidth under normal incidence of the y-polarized wave (the same as the x-polarized wave). With the increase of the radius of the metallic disk, the bandwidth decreases and the first resonant frequency shows a blue shift while there is a large red shift on the second and third frequencies. When r3=1.5mm, the first two resonant frequencies come into one frequency. The above analysis is also available for the oblique incidence under $\theta=40°$ as shown in Fig.5(b), except that there are only two resonant frequencies under oblique incidence even if r3 varies from 1.1 to 1.6mm. Additionally, we calculate the minimum PCR between the first resonant frequency and the last resonant frequency from the simulated results. As shown in Fig.6, when the radius of the metallic disk r3 becomes longer, the minimum PCR decreases gradually under normal incidence, but it rises gradually when r3 is lower than 1.6mm under $\theta=40°$. Especially, the minimum values of PCR are all higher than 90% under both $\theta=0°$ and $\theta=40°$ when r3 is 1.5mm.

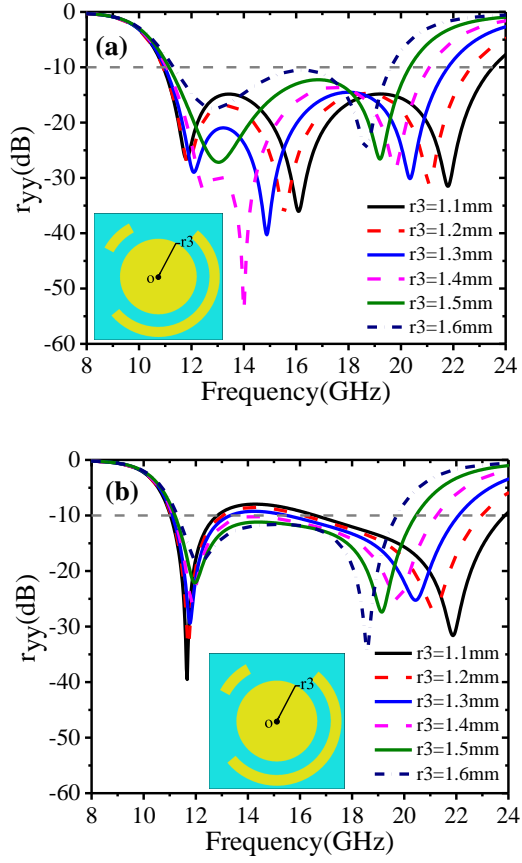

Fig.5. Simulated co-polarization reflection coefficients with r3 varying from 1.1 to 1.6 mm;(a) incident angle $\theta=0°$; (b) $\theta=40°$.

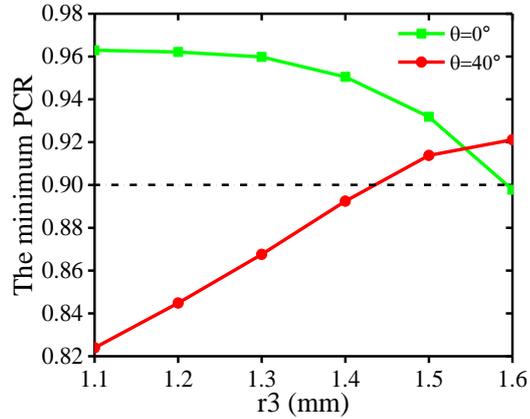

Fig.6. Simulated minimum PCR between the first resonant frequency and the last resonant frequency with r3 varying from 1.1 to 1.6 mm.

### 3.2 Incident angle insensitivity analysis

The most important criteria for obtaining a very weak angular dependence is to avoid coupling to propagating first negative diffraction orders, which is related to the unit cell periodicity *p* of the polarization conversion metasurface according to the Bragg's law [46]. In addition, the thickness of the substrate also exerts a strong influence on the angle independence [33]. As shown in Fig.7(a), the electromagnetic waves run back and forth in the dielectric substrates. The reflected coefficient on the boundary between the PET and the PVC layer is $\Gamma = (\sqrt{\varepsilon_2} - \sqrt{\varepsilon_1})/(\sqrt{\varepsilon_2} + \sqrt{\varepsilon_1}) = 0.012$, where $\varepsilon_1$ and $\varepsilon_2$ represent the real parts of dielectric constants of the PET and PVC materials, respectively. Due to the thin thickness d1 of the PET and the small value of $\Gamma$, the two layers can be equivalent to one PVC layer with the thickness d1+d2 in Fig. 7(b). The additional phase due to the oblique incidence is defined as $\Delta\varphi_{add}$,

$$\Delta\varphi_{add} = \varphi_{obl} - \varphi_{nor} = 2\sqrt{\varepsilon_2}k_0(d_1+d_2)(1/\cos\gamma - 1)$$
$$= 2\sqrt{\varepsilon_2}k_0(d_1+d_2)(1/\sqrt{(1-(\sin^2\theta/\varepsilon_2))} - 1) \quad (8)$$

where $\varphi_{obl}$ and $\varphi_{nor}$ are the phases of the incident wave in the dielectric layer under oblique incidence and normal incidence, respectively, $\gamma$ represents the refraction angle, $k_0$ is the wave number in the air. When other parameters are the same, the thinner thickness leads to the less additional phase, which would be benefit to keep better polarization conversion ratio under oblique incidence.

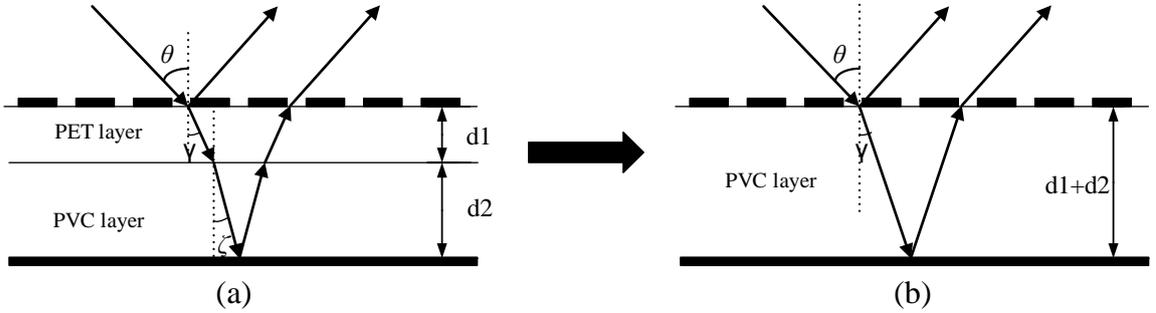

Fig. 7 Diagrammatic sketch of EM wave propagation in (a)two dielectric plates (b) a dielectric plate for oblique incidence.

Our designed polarizer has optimized the size and the thickness of the unit cell, here we make a comparison with some other reflection linear polarization converters in the Table I. In Table I, BW means the operating absolute bandwidth(PCR>90%) of the metasurface. The parameters *p* and *t* are the unit cell periodicity of the polarization conversion metasurface and the dielectric substrate thickness of the metasurface, respectively. $\Delta\varphi_{add}$(max) is defined as the maximum additional phase under the incident angle of 45 ° and $\lambda_0$ is the wavelength of the incident wave in the air. The incident angle dependence can be reduced by our designed polarization converter despite the fact that the bandwidth is reduced at the same time.

TABLE I. Comparison with other wideband polarization conversion metasurfaces.

|  | $p$(mm) | $t$(mm) | BW(GHz) | $p/\lambda_0$(max) | $\Delta\varphi_{add}$(max) |
|---|---|---|---|---|---|
| Ref.[29] | 12.5 | 4.0 | 5.2~12.1 | 0.504 | 15.58° |
| Ref.[32] | 10 | 3 | 6.84~16 | 0.533 | 28.07° |
| Ref.[33] | 6.4 | 1.6 | 12.4-27.96 | 0.596 | 28.87° |
| Present study | 5 | 2.12 | 11.3~20.2 | 0.338 | 17.07° |

Then, we study the polarization conversion of the optimized structure under the oblique incidence of y-polarized EM waves. Fig.8(a) presents that the first resonance gets weaker and amplitudes of $r_{yy}$ at last resonant frequency keep below -20dB with the increase of $\theta$ from 0° to 45°. Meanwhile, Fig.8(b) shows that cross-polarization reflection coefficients are not greatly affected by the incident angle from 0° to 45°. The values of PCR are calculated from those simulated co- and cross-polarization reflection coefficients. As can be seen from Fig.9(a), even if $\theta$ is increased to 40° and 45°, the values of PCR can still keep higher than 90% from 11.2GHz to 20.3GHz and over 85% from 11.2GHz to 20.5GHz, respectively. BW keeps nearly unchanged under $\theta$ increasing from 0° and 45°. The simulated results further demonstrate that our designed polarization converter can work well with a wide range of incident angles. Furthermore, compared with the symmetric structure composed of the split-ring resonators and metallic disk resonators (together henceforth referred to as DSRRs) in [31], our proposed asymmetric structure can avoid the split of the bandwidth which is caused by the absorption, as shown in Fig.9 (a) and (b).

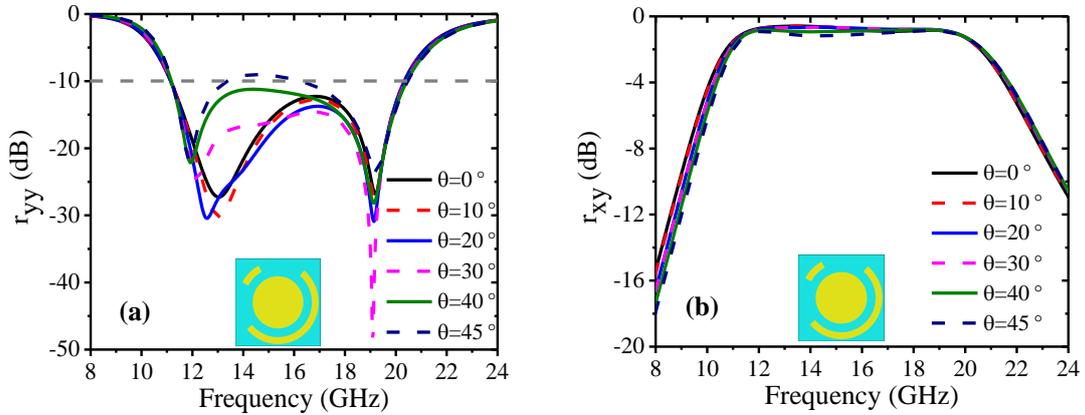

Fig.8. Simulated (a) co-polarization, (b) cross-polarization reflection coefficients of the proposed polarizer under the oblique incidence of y-polarized electromagnetic waves when $\theta$ varies from 0° to 45° (r3=1.5mm).

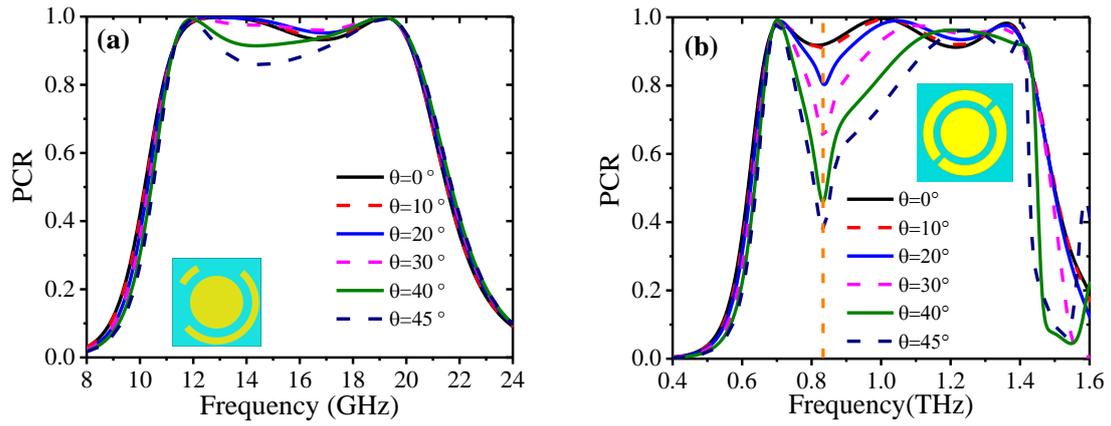

Fig.9. Simulated PCR of (a) the proposed polarizer and (b) the polarizer based on the DSRRs under the oblique incidence of y-polarized electromagnetic waves with $\theta$ from 0° to 45°(r3=1.5mm).

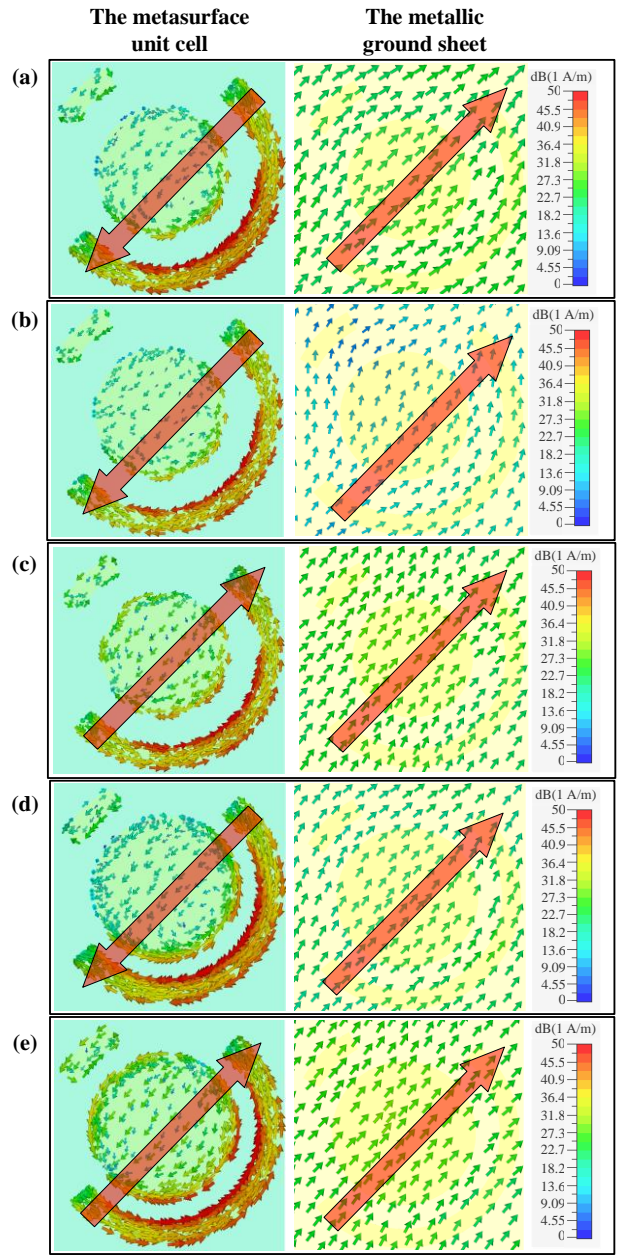

Fig.10. The current distributions on the metasurface (the left column) and metallic ground sheet (the right column) under y-polarized wave incidence when r3=1.4mm at (a) 12.5GHz, (b) 14.0GHz, (c) 19.8GHz, and when r3=1.5mm at (d) 13.0 GHz, (e) 19.2 GHz, respectively. The red arrow indicates the equivalent current direction.

**3.3 Current distributions**

From the perspective of resonances, the surface current distributions are given in Fig.10 at resonant frequencies under r3=1.4mm and r3=1.5mm (different resonant numbers as shown in Fig.5(a) and similar minimum PCR values in Fig.6) to explain the physical mechanism of the reflected linear polarization converter. There are three resonances, respectively at 12.5GHz, 14.0GHz and 19.8GHz when r3=1.4mm. The surface current distributions at these resonant frequencies on the top metallic structure and metallic ground sheet are shown in Figs.10(a)~10(c). The top metallic structure forms an equivalent surface current at 12.5GHz and 14.0GHz, which is antiparallel to the current flowing on the bottom ground plane. In this case, an equivalent circulating current is formed between the top metallic structure and the metallic ground, and magnetic resonances are created at those two resonant frequencies. On the contrary, the surface current on the top metallic structure is parallel to the current on the ground plane at 19.8GHz, which can be regarded as an electric dipole that can excite the electric resonance. Under r3=1.5mm, there is a magnetic resonance at 13.0GHz and an electric resonance at 19.2GHz in the same analytical way, as shown in Figs.10(d), (e).

We can find that the current on the metallic ground sheet at 14.0GHz in Fig.10(b) is much weaker than that at other resonant frequencies, which means that the magnetic resonance at 14.0GHz is easily broken. Moreover, with the increase of the incident angle, the magnitude of the electric field along u-axis and v-axis (y-polarized wave decomposed into the two directions) would decrease, which would lead to the weaker magnetic resonance. Thus, the second resonant frequency would disappear when the incident angle $\theta=40°$ as shown in Fig.5(a) and (b).

Furthermore, it can be seen that most of the current distributes on the long arc metallic wire and metallic disk and a little on the short arc metallic wire, which confirms long arc metallic wire and metallic disk make greater contributions to resonances.

It is also necessary to point out that the y-polarized wave (the same as the x-polarized wave) could excite all the three resonant modes(I), (II) and (III) in Fig.3(a) because the electric field of it can be decomposed into two components along u-axis and v-axis, respectively. These resonant modes would generate resonance frequencies of different positions and numbers through mutual coupling.

### 3.4 Conformal performance

In order to study the conformal performance, the designed polarizer and the pure metallic sheet are bent on convex cylindrical surfaces respectively with different curvature radiuses R=250mm and R=500mm, as shown in Fig.11(a), (b) and (c), (d). We define $\sigma_{yy}$ as the co-polarization radar cross section (RCS) and $\sigma_{xy}$ as the cross-polarization RCS under the y-polarized plane wave,

$$\sigma_{yy} = 4\pi D^2 |\mathbf{E}_y^r|^2 / |\mathbf{E}_y^i|^2 \quad and \quad \sigma_{xy} = 4\pi D^2 |\mathbf{E}_x^r|^2 / |\mathbf{E}_y^i|^2 , \qquad (9)$$

where D is the distance from the metasurfaces to the observation points, $\mathbf{E}^r$ and $\mathbf{E}^i$ are the scattered and incident electric fields, respectively, $\mathbf{E}_x$ and $\mathbf{E}_y$ are electric fields along the x-axis and y-axis respectively. RCS can be used to analyse the polarization conversion of our designed polarizer on curved surfaces with the same D based on equations (9). Fig.11(a),(c) and (b),(d) show the simulated 3-D bistatic RCS including co-polarization RCS $\sigma_{yy}$ and cross-polarization RCS $\sigma_{xy}$ under normal incidence of y-polarized EM wave with our designed polarizer and the same shaped metal cylindrical metallic sheets, respectively. For metallic curved sheet, main scattered energy distributes in the co-polarization component and little in the cross-polarization component, as can be seen in Fig.11(b) and (d). On the contrary, cross-polarization RCS is much higher than co-polarization RCS in the main scattered area for our designed polarizer under R=250mm and R=500mm in Fig.11(a) and (c), which indicates that our proposed polarizer conformal on the curved surface can achieve polarization conversion. Besides, in the column titled with cross-polarization of the Fig.11(a) and (c), when R increases from 250mm to 500mm, the energy of the scattered wave propagating along +z-axis would increase and the beamwidth of the lobe along +z-axis in the xoz plane would be narrower, which indicates that the energy of the scattered wave would be more concentrated.

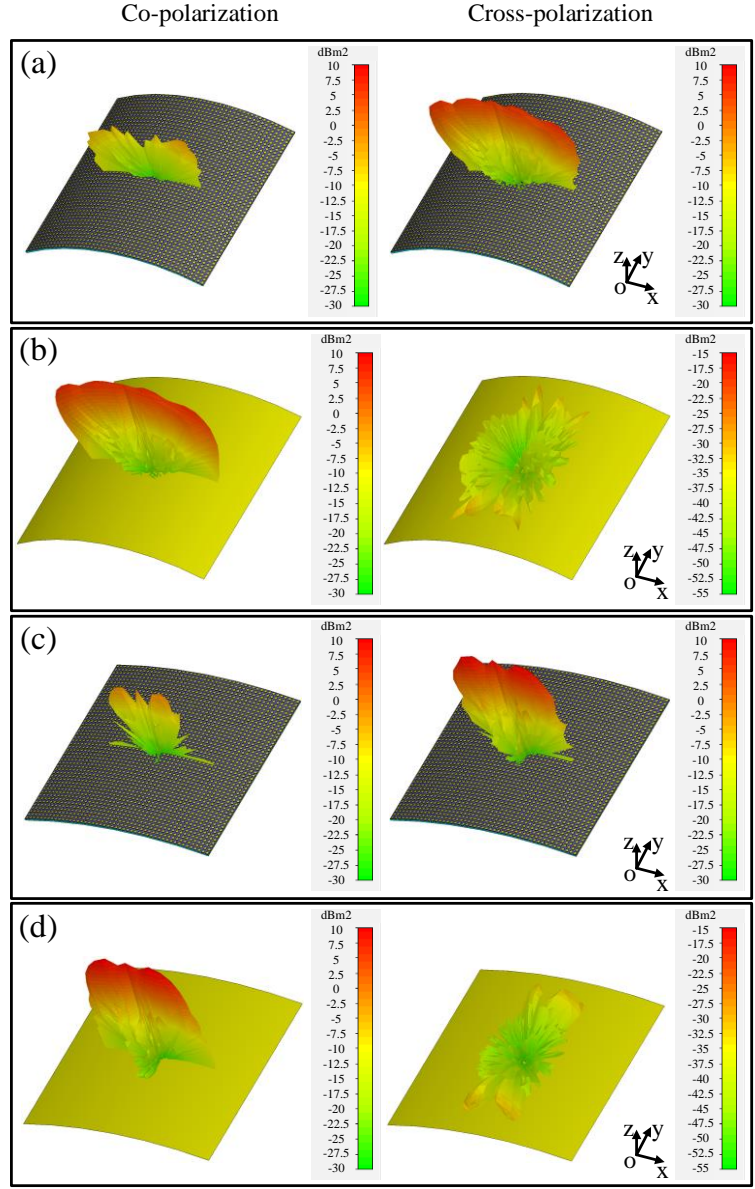

Fig.11. Bi-static 3D $\sigma_{yy}$ and $\sigma_{xy}$ of the polarizer on the convex cylindrical surface((a) and (c)),and metal convex cylindrical surface((b) and(d)), with different curvature radiuses R=250 mm((a) and (b)) and R=500 mm((c) and (d)) at 13.0GHz.

## 4. EXPERIMENTS

To experimentally show properties of the proposed polarization converter, some prototypes are fabricated on PET substrate and stuck on the PVC substrate with these parameters (d1=0.12mm, d2=2mm, w=0.4mm, r1=r2=2.4mm, r3=1.5mm, $\alpha$=190°, $\beta$=33°). Fig.12 shows the photo of the fabricated prototype and the measurement setup. The sample is placed in the front of the horn antennas surrounding by absorbing materials and the horn antennas are fixed on the two vertical arms of the rotating pedestal respectively. The centers of the sample and horn antennas are set at the same height. One horn antenna is used as the source and the other is used for receiving reflected waves to obtain the co- and cross-polarization reflective coefficients by rotating the receiving antenna by 0° and 90°. Both antennas are connected to an Agilent vector network analyzer(N5245A). The horizontal arms connected to the rotating pedestal can be rotated to obtain the reflective coefficients under different incident angles. However, the separation angle between the two antennas is set to be 5° firstly because of their overlapping positions under normal incidence. It should be noted that the metal ground sheet is also measured for the sake of normalization. Fig.13 and Fig.14 give the simulated and measured reflective coefficients, which indicate that the experiment results are in good agreement with numerical predictions. We can clearly observe two resonant frequencies when the incident angle varies from 5° to 45°. PCR is also given in Fig.15(a), which is in good agreement with the simulation. Moreover, in order to show the conformal feature of our proposed polarizer, some arc foams have been made and the sample is stuck on the different convex cylindrical surfaces of them and tested. Limited by the testing condition, the measured bistatic RCS cannot be obtained at present and we use S-parameters to evaluate the polarization conversion. The same cylindrical metal ground sheets are measured for the sake of normalization without taking influences of scattering waves brought by the cylindrical structure into consideration. Fig. 15(b) shows the calculated PCR from the measured results. The designed polarization converter can keep high polarization conversion efficiency from 11.9GHz to 20.0GHz even it is bent on convex cylindrical surfaces.

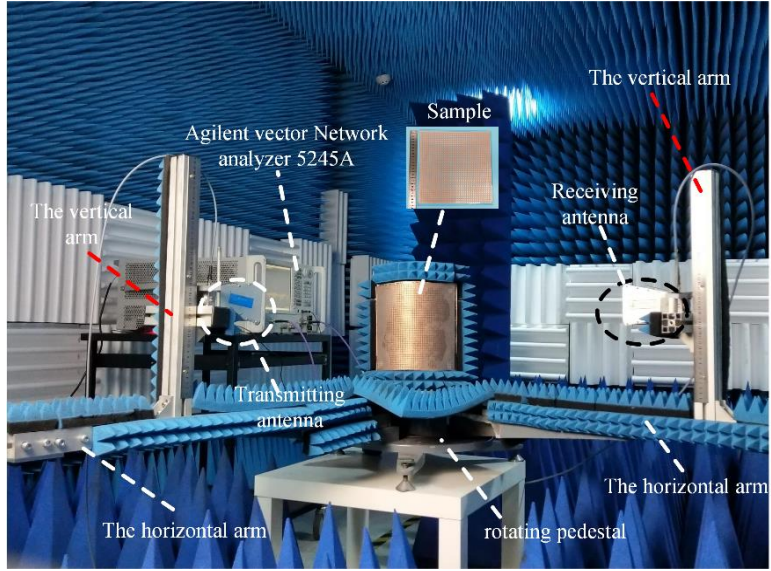

Fig.12. Fabricated metasurfaces with parameters (d1=0.12mm, d2=2mm, w=0.4mm, r1=r2=2.4mm, r3=1.5mm, α=190°, β=33°) and experiment setup.

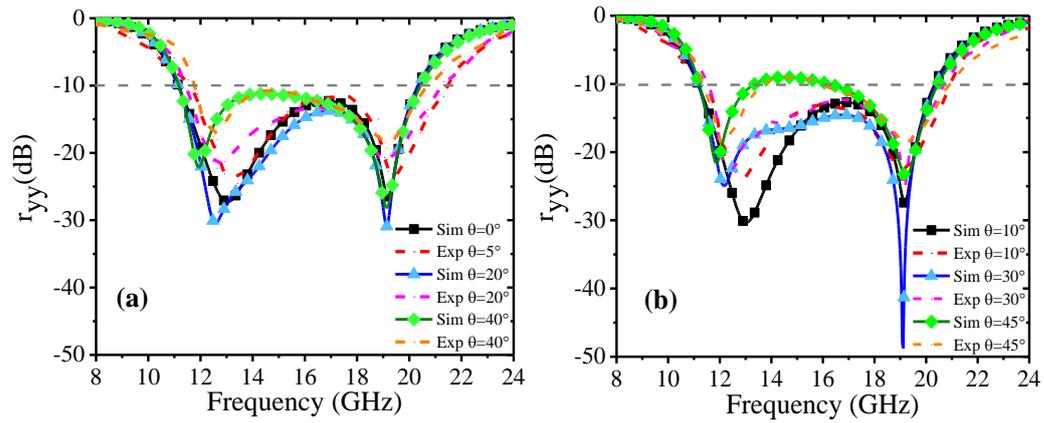

Fig. 13 The simulated and measured magnitude of $r_{yy}$ (a) with θ=0°,20°,40°;(b) with θ=10°,30°,45°.

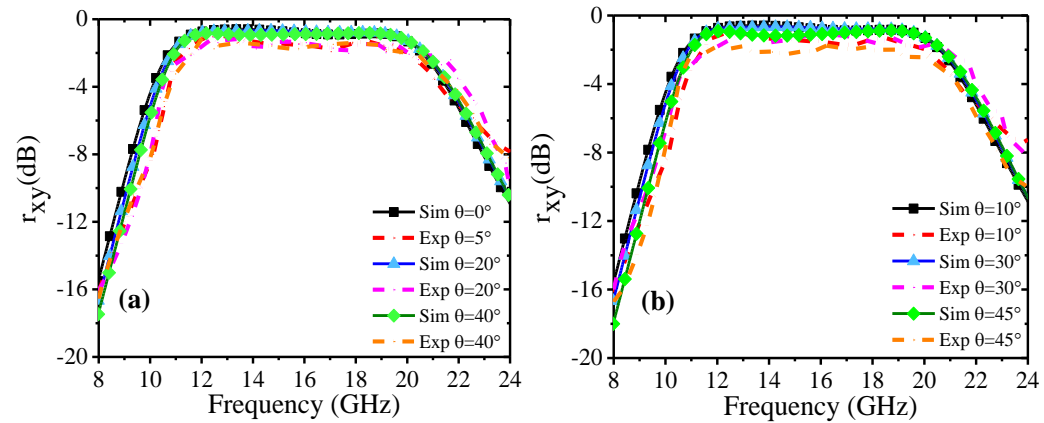

Fig. 14 The simulated and measured magnitude of $r_{xy}$ (a) with $\theta=0°,20°,40°$;(b) with $\theta=10°,30°,45°$.

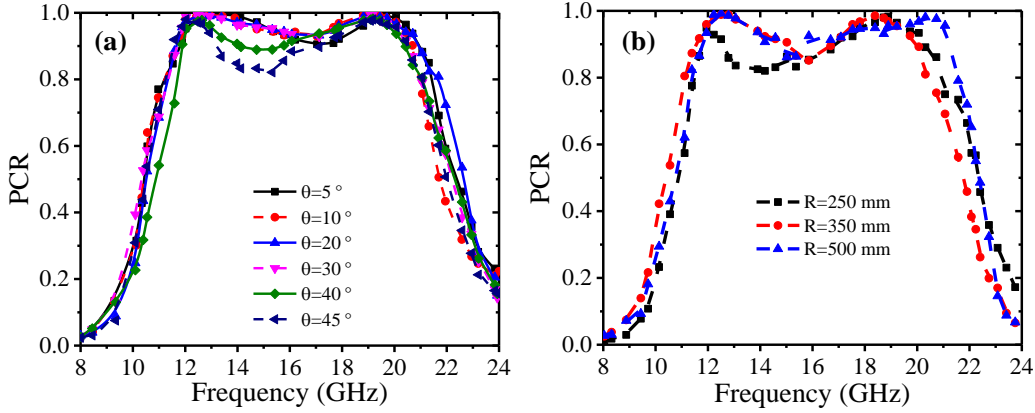

Fig.15. (a) The experimental PCR with $\theta$ varying from 5° to 45° when the sample is placed on the plane. (b) The experiment PCR when the sample is stuck on the convex cylindrical surface with R varying from 250 to 500 mm.

## 5. CONCLUSION

In this paper, we have proposed a high-efficient, broadband and wide-angle reflective polarization converter based on flexible metasurface at microwave band. The operation bandwidth has been expanded greatly because of the presence of these resonances. The physics mechanism is illustrated by simulating the reflected amplitudes and phase differences between u- and v-polarized incident EM waves. Due to the combination of the first two resonances into one and the optimized size and thickness of the unit cell, the polarization converter can keep high-efficient polarization conversion over a broadband frequency range when the angle of incident electromagnetic waves varies from 0° to 45°. The surface current distributions at resonant frequencies are also given to further analysis the physical mechanism of the polarizer. Both simulation and experiment results have shown that the values of PCR can keep over 90% from 11.2GHz to 20.3GHz when incidence angle changes from 0° to 40°. Additionally, the sample stuck on convex cylindrical surfaces with different curvature radiuses R is tested and the tested results show good polarization conversion and flexibility of our designed polarizer. Compared with previous researches, this one can not only have wide bandwidth, high efficiency and a weak angular dependence, but also be conformal with the surfaces of the automobiles, airplanes, steamships and so on. The designed structure has great potential application values in the microwave, terahertz and optic regimes.